\documentclass[a4paper,12pt]{article}
\usepackage{epsfig}
\usepackage[dvips,usenames]{color}
\usepackage{axodraw}
\usepackage{graphicx}

\newlength{\dinwidth}
\newlength{\dinmargin}
\begin{document}
\title{Testing the chirality of b to u current with  $B^{0}
\to \rho^{-} \ell^{+} \nu$ }
\bigskip
\author{Yadong Yang\footnote{ Corresponding
author. E-mail address: yangyd@henannu.edu.cn},~Hongjun Hao and
Fang Su
\\
{ \small \it Department of Physics, Henan Normal University,
Xinxiang, Henan 453007,  P.R. China}\\
}
\bigskip\bigskip
\maketitle

\begin{abstract}
\noindent

We study the effects of a modest V+A admixture strength  in $b\to
u$ current in the decay $B \rightarrow\rho l \nu$. We have shown
that the decay rate, lepton forward-backward distribution
asymmetry $A_{FB}$ and polarization ratio are sensitive to the
admixture. Future experimental studies of the decay at BaBar and
BELLE could  clarify the chirality of $b\to u$ current and might
reveal hints for New Physics with right-handed quark currents.

\end{abstract}

\bigskip\bigskip

{\bf PACS Numbers: 12.60.Cn, 12.15.Ji, 13.20.He}

\newpage
With the running of BaBar and Belle B factories, many rare B
decays could be well studied and provide tests for the Standard
Model(SM). Potentially it will open windows for New Physics beyond
the SM, since clear differences between the  measurements of
observables in B decays and  the SM expectations will be  indirect
evidences for New Physics existence.

 In the SM, the weak charge currents are left-handed(LH).
Therefore,  derivations from pure V-A chirality of weak charge
currents would be important probes for New Physics. Years ago, the
speculation of right-handed chirality $b\to c$ current had been
put forward by K\"orner and Schuler \cite{kroner},  then by Gronau
and Wakaizumi\cite{Gronau}, Hou and Wyler\cite{hou}. Since then,
the problem has been studied by many authors\cite{studies,
studies2}. With the experiment measurements by \cite{CLEO1}and
L3\cite{L3}, the dominant right-handed $b\to c$ current was ruled
out. However, a modest right-handed admixture in $b\to c$ current
is still allowed. Moreover, as shown  by Voloshin\cite{Vol}, it
may help to resolve  the baffling semileptonic branching ratio and
the charm-deficit problems in B decays\cite{bigi}. Such a
situation was reexamined comprehensively by Rizzo\cite{Rizzo}
later on. Compared to the extensive studies of the chirality of
$b\to c $, the relevant studies about a modest right-handed
admixture in $b\to u$ current are quite few. We note that
Wakaizumi\cite{Wakai} studied it using the Left-Right Symmetric
Model. The main reason for the situation may be due to the fact
that $b\to u$ transitions are much rarer than $b\to c$. However,
many rare decays induced by $b\to u$  could be well studied now at
BaBar and Belle factories. For example, using a sample of $3\times
10^6  \Upsilon(4S)\to B{\bar B}$ events, CLEO \cite{CLEO2} has
given a quite accurate  measurement of $B\to \rho l\nu$ decay. It
is interesting to note that  both BaBar and Belle have accumulated
more than 200M $B{\bar B}$ events by 2004. We can foresee that
observables in the decay could be well measured by BaBar and Belle
in the near future.

In this letter, we will investigate the sensitivities of lepton
angular distribution asymmetry and  $\rho$ meson polarization in
$B^0 \to \rho^- \ell^+ \nu$ decay to a modest V+A admixture in
$b\to u$ current. As shown later, these observables are very
sensitive to the admixture.

Following Voloshin\cite{Vol}, we introduce a modest V+A admixture
in  $b\to u$ coupling, while maintain the leptonic current as
purely left-handed to satisfy the tight constraints from $\mu$
decays\cite{muon}. So the effective Hamiltonian describing the
decay can be written as
\begin{eqnarray}
{\cal H}_{eff}
&=&-4\frac{G_{F}}{\sqrt{2}}V_{ub}[(\overline{u}_{L}\gamma_{\mu}b_{L})+
\xi(\overline{u}_{R}\gamma_{\mu}b_{R})](\overline{l}_L\gamma^{\mu}\nu_L)+h.c.,
\end{eqnarray}
where $G_{F}$ is the Fermi  constant, $V_{ub}$ is the KM matrix
element for the $b\rightarrow u$ transition, and $\xi$ denotes the
relative strength of the RH admixture to the LH  $b \to u$
coupling. As we do not discuss CP violation, we will treat $\xi$
as a real parameter.

Neglecting lepton masses, the double differential decay rate in
$q^2$ and $\cos\theta$ is given in terms of three
$q^{2}$-dependent helicity amplitudes ${H}_{0, \pm}$, where the
subscripts refer to the helicity of the $\rho$-meson\cite{bauer}
\begin{eqnarray}
\frac{d^{2}\Gamma(B^{0}\rightarrow\rho^{-} \ell^{+}\nu)}{dq^{2}d
\cos{\theta}} =\frac{G_{F}^{2}{|V_{ub}|}^{2}}{256\pi^{3}}
K_{\rho}\frac{q^{2}}{m^{2}_{B}}\left[{(1-\cos{\theta})}^{2}{|
H_{+}|}^{2}+{(1+\cos{\theta})}^{2}
{|H_{-}|}^{2}+2{\sin}^{2}\theta{|H_{0}|}^{2}\right].
\end{eqnarray}
Where $q$ is the momentum of  lepton pair, $\theta$ is the angle
of the lepton measured  with respect to the $\rho$-direction in
the $l\nu$ pair rest frame; $K_{\rho}$ is the absolute value of
$\rho$ meson three-momentum in the B rest frame,
\begin{eqnarray}
K_{\rho}&=&\frac{1}{2m_{B}}{[{({m_{B}}^{2}-{m_{\rho}}^{2}-q^{2})}^{2}
-4{m_{\rho}}^{2}q^{2}]}^{\frac{1}{2}}.
\end{eqnarray}

The helicity amplitudes $H_{0,+,-}$ are related to the
Lorentz-invariant form factors as
\begin{eqnarray}
H_{\pm}(q^{2})& = &(m_{B}+m_{\rho})(1-\xi)A_{1}(q^{2})\mp
\frac{2m_{B}K_{\rho}}{m_{B}+m_{\rho}}(1+\xi)V(q^{2}),\\
H_{0}(q^{2})& = &\frac{1-\xi}{2m_{\rho}\sqrt{q^{2}}} \left[
({m_{B}}^{2}-{m_{\rho}}^{2}-q^{2})(m_{B}+m_{\rho})A_{1}(q^{2})
-\frac{4{m_{B}}^{2}K^{2}_{\rho}}{m_{B}+m_{\rho}}A_{2}(q^{2})
\right].
\end{eqnarray}
Form factors V and $A_{1,2}$ are connected with the meson
transition amplitudes, induced by the vector
$V^{\mu}=\overline{u}\gamma^{\mu}b$ and axial-vector
$A^{\mu}=\overline{u}\gamma^{\mu}\gamma^{5}b$ quark transition
currents.

All the dynamical content of hadronic current matrix elements is
described by the above $q^{2}$-dependent form factors. The
calculation of these form factors requires non-perturbative
methods, and are sources of large theoretical uncertainties.
Theoretical approaches for calculating these form factors are
quark model, QCD sum rule, and lattice QCD. In our calculations,
we use both results from light-cone sum rule (LCSR)\cite{Ball} and
lattice QCD mode(LQCD)\cite{U} for comparison.

Experimentally, the main difficulty in observing signals from
$b\rightarrow u \ell \nu$ processes is the very large background
due to $b\rightarrow c\ell\nu$. With the lepton-energy requirement
$E_{l}>2.3$GeV \cite{CLEO2}, the $b\rightarrow c\ell\nu$
background can be sufficiently suppressed.   It is well known that
QCD sum rule is suitable for describing the low $q^{2}$ region of
the form factors while lattice QCD  for the high $q^{2}$ region.
So that, we shall calculate the decay width of $B^{0}\to \rho^-
 \ell^+  \nu$  in the phase space $ E_l >2.3GeV$ and $0<q^{2}<7
GeV^{2}$ using LCSR form factors\cite{Ball}, while in the phase
space 2.3$<E_{l}<2.6$GeV, $14<q^{2}<21 GeV^{2}$ using lattice QCD
form factors\cite{U}.

Using the upper bound of $q^2$ for a given $E_l$
\begin{equation}
q^{2}_{up}=2E_{l}\frac{m_{B}^{2}-m_{\rho}^{2}-2m_{B}E_{l}}{m_{B}-2E_{l}}
\end{equation}
and
\begin{equation} \cos\theta=
-\frac {4m_{B}E_{l}-m_{B}^{2}-q^{2}+m_{\rho}^{2}}{2K_{\rho}m_{B}},
\end{equation}
we can integrate Eq.2 over the phase spaces of our interests. Our
numerical results for partial widths in the phase spaces $0<leq
q^{2}<7GeV^{2}$ and  $14GeV^{2}<q^{2}<21GeV^{2}$ with the lepton
energy requirement $E_{\ell}>2.3$GeV are presented in Fig.1 and
Fig.2, respectively. We have used $|V_{ub}|=4.7\times
10^{-3}$\cite{HFAG}.
\begin{figure}
\begin{center}
\scalebox{0.8}{\epsfig{file=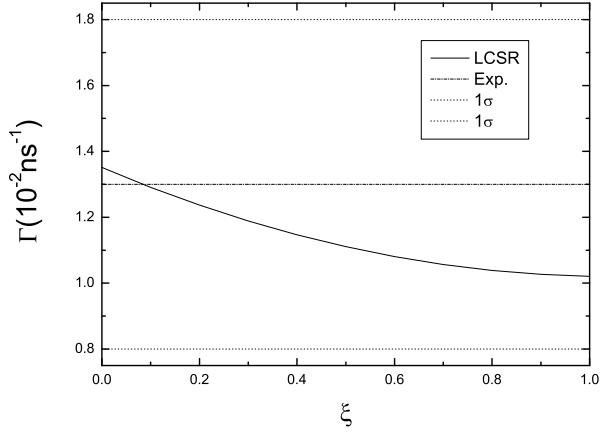}} \caption{\small The partial
rate $\Gamma$ in the phase space with $E_{l}>2.3$GeV and
0$<q^{2}<7$GeV$^{2}$  as a function of $\xi$ (solid curve ),
calculated using LCSR form factors\cite{Ball}. The horizontal
lines are the CLEO data and its $1\sigma$ error bar \cite{CLEO2} }
\end{center}
\end{figure}

From Fig.1 and Fig.2, we can find that the partial decay widths
are sensitive to the V+A admixture strength $\xi$. However, our
numerical calculations must use $V_{ub}$ as input, which is poorly
known at present. Given accurate measurements of $\Gamma(B\to
\rho\ell\nu)$ and  $\Gamma(B\to X_{u}\ell\nu)$  available from
BaBar and Belle in the near future, we still may not get a
reliable conclusion about the chirality of $b\to u$ current,
because the effects of the V+A admixture on the decay widths could
be lumped by a redefinition of effective $V_{ub}$. We may have to
resort to observables which are $V_{ub}$ independent. To this end,
we will study lepton angular asymmetries and polarization ratios
in the decay, which are free of our knowledge of $V_{ub}$.

\begin{figure}
\begin{center}
\scalebox{0.8}{\epsfig{file=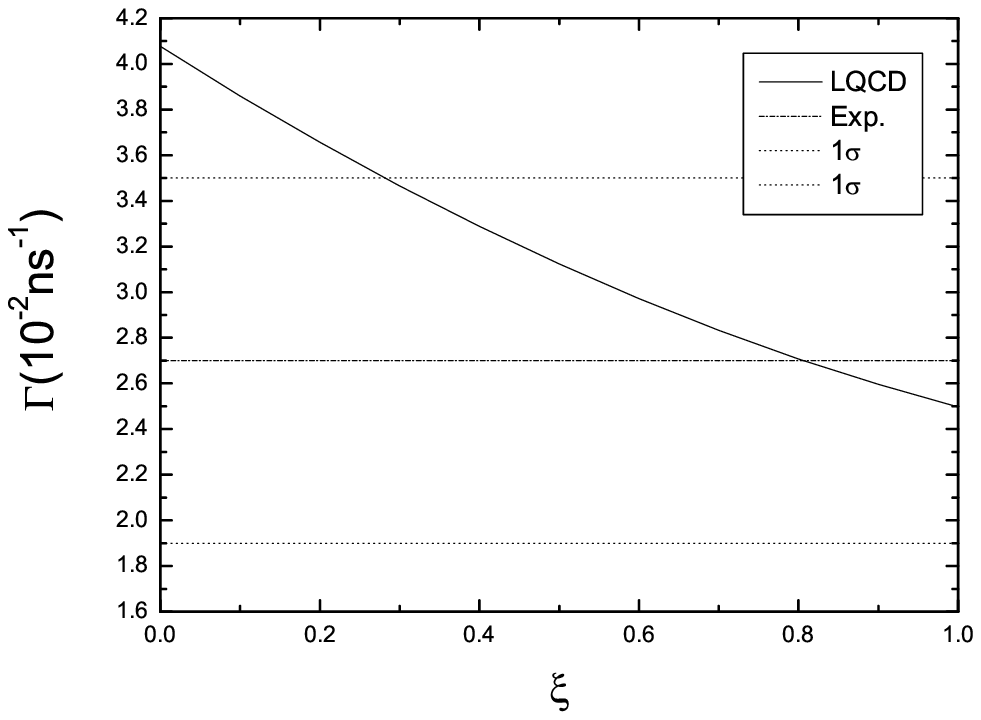}} \caption{\small The partial
rate $\Gamma$ in the phase-space with $E_{l}>2.3$GeV and
$14<q^{2}<21GeV^{2}$  as a function of $\xi$ (solid curve ),
calculated using lattice QCD form factors\cite{U}. The horizontal
lines are the CLEO data and its $1\sigma$ error bar \cite{CLEO2} }
\end{center}
\end{figure}

The forward-backward asymmetry ($A_{FB}$) of the lepton
distribution  is defined as\cite{kroner}
\begin{eqnarray}
A_{FB}=\frac{\int
 dq^{2}\,[\frac{d\Gamma}{dq^{2}}(\theta>\frac{\pi}{2})-
\frac{d\Gamma}{dq^{2}}(\theta<\frac{\pi}{2})]}{\int
dq^{2}\,[\frac{d\Gamma}{dq^{2}}(\theta>\frac{\pi}{2})+
\frac{d\Gamma}{dq^{2}}(\theta<\frac{\pi}{2})]}.
\end{eqnarray}

Another interesting observable is the polarization ratio
$\frac{\Gamma_{L}}{\Gamma_{T}}$, where $\Gamma_{L,T}$ are the
rates for producing longitudinally and transversely polarized
$\rho$ meson in the decay respectively. The polarization ratio can
be written as \cite{Neubert}
\begin{eqnarray}
\frac{\Gamma_{L}}{\Gamma_{T}}&=&\frac{\int
 dq^{2}\,2z_{0}(1-z_{0}^{2}/3)K_{\rho}q^{2}H_{0}^{2}}{\int
dq^{2}\,z_{0}(1+{z_{0}}^{2}/3)K_{\rho}q^{2}({H_{+}}^{2}+{H_{-}}^{2})},
\end{eqnarray}
where
\begin{equation} z_{0}=min[1,-\frac
{4m_{B}E^{cut}-m_{B}^{2}-q^{2}+m_{\rho}^{2}}{2K_{\rho}m_{B}}].
\end{equation}

Since leptons energy cut-off is always required in experiment
measurement and the lepton energy $E_{\ell}$ is a function of the
decay angle $\theta$, the measured values of the above observables
will be affected by $E_{cut}$. The requirement $E_{\ell}\geq
E_{cut}$ introduces a limit in the experimentally accessible
angular range $-1\leq\cos\theta\leq z_{0}$. The range is not
symmetric.
 However we need to use a
symmetric cut with respected to $\cos\theta$, i.e., $
-z_{0}\leq\cos\theta\leq z_{0}$, in order not to affect the
contributions to different helicity amplitudes in a different
way\cite{kroner, Neubert}. Since CLEO\cite{c} and BABAR\cite{b}
used $E_{cut}=2.0$GeV in their recent measurement of the rate of
$B \rightarrow\rho^{-} \ell^{+}\nu$, we will calculate the
asymmetries using the same cutoff. For comparison, we also
calculate the asymmetries  without the energy cut.

Our numerical results for $A_{FB}$ and $\Gamma_{L}/\Gamma_{T}$ are
presented in Fig.3 and Fig.4, respectively.  The solid lines are
the results with energy cut $E_{cut}=$2.0GeV and the dashed lines
for the results without energy cut.

\begin{figure}
\begin{center}
\scalebox{0.9}{\epsfig{file=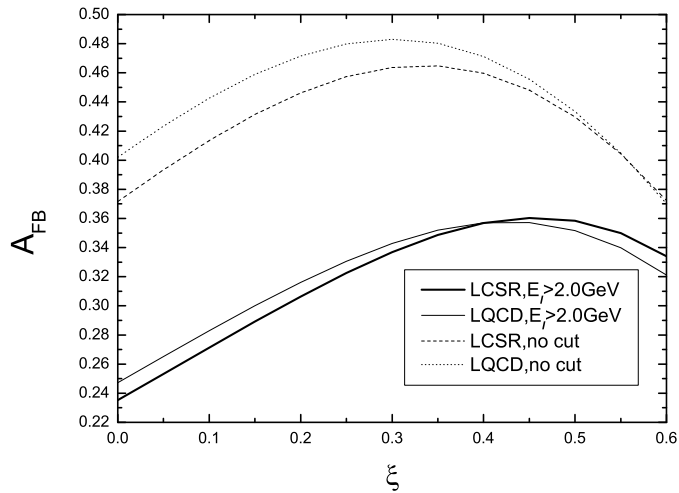}} \caption{\small The
forward-backward asymmetry $A_{FB}$ as a function of $\xi$. The
solid(dashed) curve corresponds to calculation with(without)
lepton energy cut $E_{cut}=2.0GeV$.}
\end{center}
\end{figure}

\begin{figure}
\begin{center}
\scalebox{0.9}{\epsfig{file=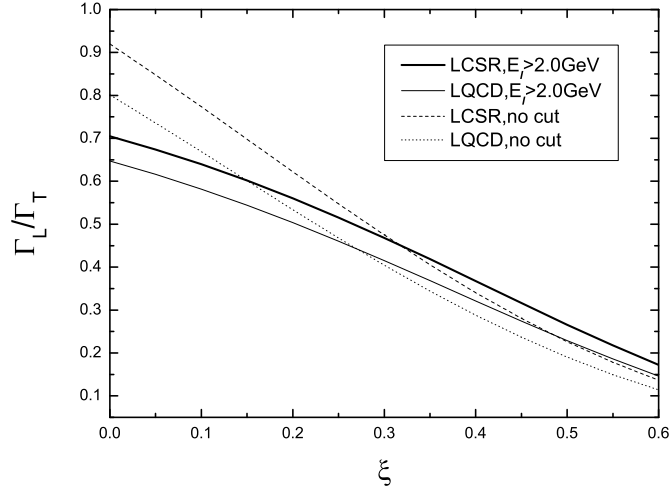}} \caption{\small The
polarization ratio $\frac{\Gamma_{L}}{\Gamma_{T}}$ as a function
of $\xi$. The solid(dashed) curve corresponds to calculation
with(without) lepton energy cut  $E_{cut}=2.0GeV$.}
\end{center}
\end{figure}

As shown in Fig.3 and 4, the $A_{FB}$ and $\Gamma_{L}/\Gamma_{T}$
are sensitive to the V+A admixture in $b\to u$ current. When
lepton energy cut required, magnitudes of $A_{FB}$ is reduced
about by half, while $\Gamma_{L}/\Gamma_{T}$ is reduced not so
much. It is interesting to note that LQCD\cite{U} and
LCSR\cite{Ball} form factors give very similar numerical results
when the lepton energy cut is put. For a modest V+A admixture
$\xi=0.2$, $A_{FB}$ will be enhanced by $17\%$,  and
$\Gamma_{L}/\Gamma_{T}$ will be reduced by $20\%$, which might be
testable  in the near future measurements at BaBar, Belle and
LHCb.

In past years, there have been considerable progresses in both
theoretical calculations of form-factors\cite{Ball} and
experimental studies of $B\to \rho l\nu$\cite{c, b}. The progress
will surely advance steadily to a higher precision stage of
calculating and measurement of the observables studied in this
paper. We note  that the recent measurement of  $Br(B\to \rho\ell
\nu)$ has been preformed by BaBar\cite{b} with $\mathbf{just}$ 55M
$B{\bar B}$ events. However, BaBar had accumulated more than 250M
$B{\bar B}$ events by 2004. Moreover, BELLE\cite{belle} has
recorded data sample as large as 375M recently. So that, refined
studies of the decay $B\to \rho\ell \nu$ could be performed by
both BaBar and BELLE to clarify the chirality of $b\to u$ weak
current, which might reveal hints for New Physics.

In summary, we have studies the effects of a modest V+A admixture
in $b\to u$ current in the decay $B\to \rho^- \ell^+ \nu $. We
have shown that the decay rate, lepton forward-backward
distribution asymmetry and $\rho$ meson longitudinal to transverse
polarizations ratio are quite sensitive to the admixture. The last
two observables, which are free of our knowledge of $V_{ub}$, are
very suitable for probing New Physics with right-handed current.
Further experimental studies of the decay at BaBar and BELLE are
urged to clarify the chirality of $b\to u$ current.

This work is supported  by NFSC under contract No.10305003, Henan
Provincial Foundation for Prominent Young Scientists under
contract No.0312001700 and in part by the project sponsored by SRF
for ROCS, SEM.

\end{document}